\documentclass[fdp,fleqn]{w-art}
\usepackage{times}
\usepackage{w-thm}
\usepackage{rotating}
\usepackage{amsmath}
\usepackage{amssymb}
\usepackage[]{graphicx}
\newcommand{\sgn}{{\rm sgn}}

\def\Z{\mathbb{Z}}

\def\ov{\overline}
\def\N{\mathbf{N}}
\def\Sym{\mathbf{Sym}}
\def\Anti{\mathbf{Anti}}
\def\Adj{\mathbf{Adj}}
\def\ov{\overline}
\def\1{{\bf 1}}
\def\2{{\bf 2}}
\def\3{{\bf 3}}
\def\4{{\bf 4}}
\def\6{{\bf 6}}
\def\OR{\Omega\mathcal{R}}

\def\targ#1#2{\genfrac{[}{]}{0pt}{}{#1}{#2}}

\begin{document}
\Volume{??}
\Issue{1}
\Month{01}
\Year{2012}
\pagespan{1}{6}
\keywords{Intersecting D-branes, Standard Model vacua, Yukawa couplings. \hspace{16mm} MZ-TH/12-08}

\title[Field theory on D6-branes: Yukawas and masses]{Towards the field theory of the Standard Model on fractional D6-branes on $T^6/\Z_6'$: Yukawa couplings and masses}

\author[G. Honecker]{Gabriele Honecker\inst{1,}%
  \footnote{Corresponding author. E-mail:~\textsf{Gabriele.Honecker@uni-mainz.de}.
}}
\address[\inst{1}]{Institut f\"ur Physik  (WA THEP), Johannes-Gutenberg-Universit\"at, D-55099 Mainz, Germany}
\author[J. Vanhoof]{Joris Vanhoof\inst{2,3}\footnote{Aspirant FWO.}}
\address[\inst{2}]{Theoretische Natuurkunde, Vrije Universiteit Brussel and The International Solvay Institutes, Pleinlaan 2, B-1050 Brussels, Belgium}
\address[\inst{3}]{Institute of Theoretical Physics, K.U.Leuven, Celestijnenlaan 200D, B-3001 Leuven, Belgium}
\begin{abstract}
We present the perturbative Yukawa couplings of the Standard Model on fractional intersecting
D6-branes on $T^6/\Z_6'$ and discuss two mechanisms of creating
mass terms for the vector-like particles in the matter spectrum,  
through perturbative three-point couplings and through continuous D6-brane displacements.
\end{abstract}
\maketitle

\section{Introduction}

While the massless spectra of D6-brane models in type IIA string theory have been discussed in a variety of cases, see e.g. the 
statistics for various orbifold backgrounds in~\cite{Gmeiner:2005vz,Gmeiner:2007we,Gmeiner:2008xq}, 
the inspection of the low-energy field theory limit has to a large extent 
focussed on the gauge couplings, see e.g.~\cite{Lust:2003ky,Akerblom:2007np,Blumenhagen:2007ip}
and~\cite{Gmeiner:2009fb} for a discussion within the string landscape.
For interaction terms such as Yukawa couplings, exact string theoretic results are only known on the six-torus, e.g.~\cite{Cremades:2003qj,Cremades:2004wa},
and its $T^6/\Z_2 \times \Z_2$ orbifold without discrete torsion, and consequently a treatment of Standard Model (SM) vacua in globally consistent supersymmetric D6-brane models on 
$T^6/\Z_6'$~\cite{Bailin:2006zf,Bailin:2007va,Gmeiner:2007zz,Bailin:2008xx,Gmeiner:2008xq,Gmeiner:2009fb}
or other $T^6/\Z_{2N}$~\cite{Blumenhagen:2002gw,Honecker:2004kb,Honecker:2004np} and $T^6/\Z_2 \times \Z_{2M}$ backgrounds with discrete 
torsion~\cite{Blumenhagen:2005tn,Forste:2010gw} has not been performed.
To go one step further in the understanding of the field theory of  intersecting {\it fractional} D6-brane SM vacua, 
we assume here that the formula for Yukawa couplings on the six-torus directly carries over to its orbifolds and extends to the new case with chiral matter at one vanishing angle,
i.e. `triangular' worldsheets with an acute angle zero. 
Our discussion focusses on the dominant worldsheet contribution to each three-point coupling.

\subsection{Geometry of a D6-brane Standard Model vacuum with `hiddden' $USp(6)$}

\begin{figure}[ht!]
\begin{center}
\includegraphics[height=32mm]{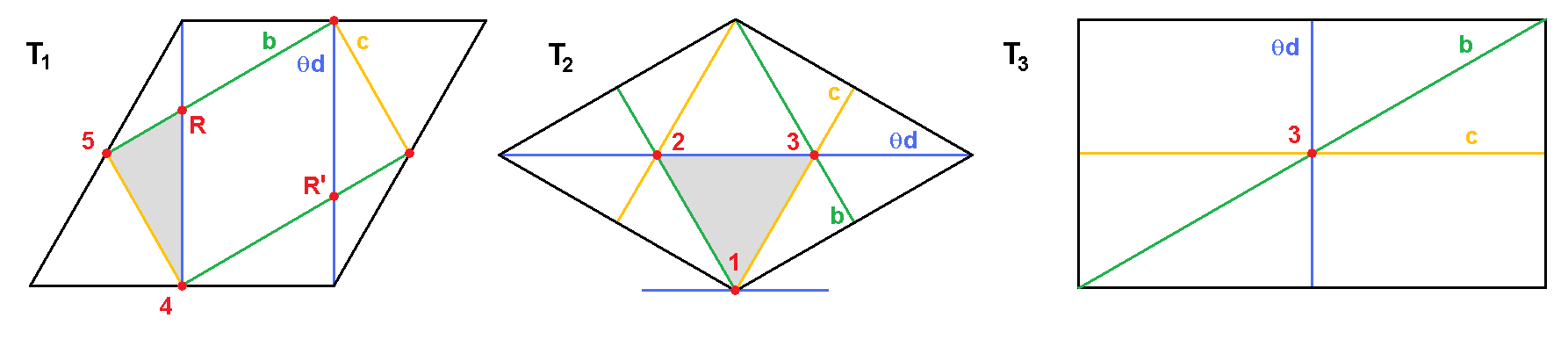}
\end{center}
\caption{The D6-branes $b$ (green), $c$ (yellow) and $(\theta d)$ (blue) on the {\bf ABa} lattice on $T^6/\Z_6'$.
Family replication in the lepton sector is due to the intersections 1,2,3 on the two-torus $T^2_{(2)}$ without $\Z_2$ action. 
The two pairs (4, $R \stackrel{\Z_2}{\leftrightarrow} R'$) and (4,5) of $\Z_2$ invariant intersection orbits along $T^2_{(1)}$ double the 
respective multiplicities of left-handed  leptons $L_i$ at $b(\theta d)$ intersections and of Higgses $H_k$ at $bc$ intersections.}
\label{fig:Leptons}
\end{figure}

The D6-brane configuration of a particular SM vacuum on the {\bf ABa} lattice orientation of $T^6/(\Z_6' \times \OR)$,
which is generated by $\theta:z_k \rightarrow e^{2\pi i v_k} z_k$ with $\vec{v}=\frac{1}{6}(1,2,-3)$ and ${\cal R}:z_k \rightarrow \ov{z}_k$, 
is given in table~\ref{RepresentationsSMOnBranes}.  
\begin{table}[ht!]
\tabsidecaption
 \setlength{\tabcolsep}{2pt}
   \begin{tabular}{|c||c|c|c|c|c|}      \hline
\multicolumn{6}{|c|}{\text{\bf D6-branes for the Standard Model with $USp(6)_h$ on $T^6/\Z_6'$ }}
\\ \hline\hline
      D6-brane & \begin{sideways} label \end{sideways}
      & \!\!\!\begin{tabular}{c}angle on\\($T_{(1)}^2$,$T_{(2)}^2$,$T_{(3)}^2$)\\ w.r.t. $\OR$-plane\\\end{tabular}\!\!\!
      & \begin{tabular}{c}displacement \\$(\sigma^1;\sigma^3)$ on \\($T_{(1)}^2$,$T_{(3)}^2$)\\\end{tabular} 
      & \begin{tabular}{c}orientation\\ on $T^2_{(2)}$ vs. \\$\OR$-plane \end{tabular}
      & \begin{tabular}{c}gauge\\group\\\end{tabular} \\ \hline\hline
      Baryonic & $a$ & $\pi\left(-\frac{1}{3},-\frac{1}{6},\frac{1}{2}\right)$ & $(1;1)$  &             & $U(3)_a$   \\
      Left     & $b$ & $\pi\left(\frac{1}{6},-\frac{1}{3},\frac{1}{6}\right)$  & $(1;0)$  &             & $U(2)_b$   \\
      Right    & $c$ & $\pi\left(-\frac{1}{3},\frac{1}{3},0\right)$      & $(1;1)$ & $\parallel$ & $USp(2)_c$  \\
      Leptonic & $d$ & $\pi\left(\frac{1}{6},\frac{1}{3},-\frac{1}{2}\right)$  & $(0;1)$  &             & $U(1)_d$   \\
      Hidden   & $h$ & $\pi\left(-\frac{1}{3},-\frac{1}{6},\frac{1}{2}\right)$ & $(0;0)$  & $\perp$     & $USp(6)_h$ \\
      \hline
   \end{tabular}
\caption{Five supersymmetric stacks of D6-branes, which cancel all RR tadpoles on the {\bf ABa} lattice of $T^6/\Z_6'$
with SM spectrum and a `hidden' $USp(6)_h$ gauge factor. The enhancements $U(N) \to USp(2N)$
arise for D6-branes parallel or perpendicular to the O6-planes. Only in the first case, the enhancement can be 
cancelled by a continuous displacement $\sigma^2$.
}
\label{RepresentationsSMOnBranes}
\end{table}
The resulting massless open string spectrum consists of a `chiral' part computed from intersection numbers (i.e. including
the anomalous $U(1)_b \subset U(2)_b$ charge of otherwise vector-like matter), which
contains the SM quarks and leptons and nine Higgs generations~\cite{Gmeiner:2008xq,Gmeiner:2009fb},
\begin{eqnarray}\hspace{-10mm}
\left[C \right]\hspace{-3mm}&=&\hspace{-3mm} 3\times\left[(\3,\2)_{\frac{1}{6}}+(\ov{\3},\1)_{\frac{1}{3}}+(\ov{\3},\1)_{\frac{-2}{3}}+(\1,\1)_{1}+(\1,\1)_{0}
+2\times(\1,\2)_{\frac{-1}{2}}+(\1,\2)_{\frac{1}{2}}\right]  
+9\times\left[(\1,\ov{\2})_{\frac{-1}{2}}+(\1,\ov{\2})_{\frac{1}{2}}\right] \nonumber
\\
&\equiv& 3\times\left[Q_{L}+d_{R}+u_{R}+e_{R}+\nu_{R}+2\times L+\ov{L}\right]+9\times\left[H_{d}+H_{u}\right] 
, \label{FullSpectrum1} 
\end{eqnarray}
where we already used the breaking $USp(2)_c \to U(1)_c$ by a continuous displacement $\sigma^2_c$. 
Besides the lower case index of the hyper charge, $Q_Y=\frac{Q_a}{6}+\frac{Q_c}{2}+\frac{Q_d}{2}$, also the baryon minus lepton number symmetry,
$Q_{B-L}=\frac{Q_a}{3} + Q_d$, is gauged and anomaly-free in this model.  
The `vector-like' spectrum from vanishing intersection numbers consists of two parts, the first with SM charges only,  
\begin{eqnarray}
[V]&=&  \Bigl[(\3,\2)_{\frac{1}{6}}+3\times(\ov{\3},\1)_{\frac{1}{3}}  +3\times(\ov{\3},\1)_{\frac{-2}{3}} 
+ 3\times(\ov{\3}_{\Anti},\1)_{\frac{1}{3}} +6\times(\1,\3_{\Sym})_{0}+c.c.\Bigr] \nonumber \\
&& +\Bigl[1_{m}\times(\1,\ov{\2})_{\frac{-1}{2}}+1_{m}\times(\1,\ov{\2})_{\frac{1}{2}}+2_{m}\times(\1,\2)_{\frac{-1}{2}}+1_{m}\times(\1,\2)_{\frac{1}{2}} 
 +4_{m}\times(\1,\1_{\Anti})_{0}+c.c.\Bigr] 
\nonumber  \\
&&+\Bigl[2_{m}\times(\1,\1)_{1}+1_{m}\times(\1,\1)_{0}+c.c.\Bigr] 
+ 2\times({\bf 8}_{\Adj},\1)_{0}+10\times(\1,\3_{\Adj})_{0}+26\times(\1,\1)_{0}
\label{FullSpectrum2}
,
\end{eqnarray}
and the second consisting of massless states with charges under the `hidden' gauge group $USp(6)_h$,
\begin{eqnarray}\hspace{-2mm}
[H]&=&2\times(\1,\1;{\bf 15}_{\Anti})_{0}+(\1,\2;{\bf 6})_0 +(\1,\ov{\2};{\bf 6})_0 
+2\times\Bigl[(\1,\1;{\bf 6})_{\frac{1}{2}}+(\1,\1;{\bf 6})_{\frac{-1}{2}}\Bigr] 
.
\label{FullSpectrum3} 
\end{eqnarray}
To investigate the interactions among the massless open string states, it is necessary to know their origin from
a given intersection sector $x(\theta^k y)_{k \in \{0,1,2\}}$ of a D6-brane $x$ with the orbifold images $(\theta^k y)$ of $y$, which can e.g. be computed using the beta
function coefficients of gauge threshold amplitudes~\cite{Gmeiner:2009fb}. In addition, the localisations of the intersection points along $T^6/\Z_6'$ 
are needed. As examples, the localisations of leptons and of quarks and Higgses  are displayed in table~\ref{tab:leptons} and~\ref{tab:quarks+Higgs},
 respectively (for more details see~\cite{Honecker:2012jd}).  
\begin{figure}
\begin{minipage}{74mm}
\setfloattype{table}
 \setlength{\tabcolsep}{1pt}
   \begin{tabular}{|c|c|c|c|c|c|}      \hline
\multicolumn{6}{|c|}{\text{\bf Multiplicities and localisations of leptons}}
\\ \hline\hline
 Particle &      $x(\theta^k y)$           &  $\chi/\varphi_{x(\theta^k y)}$   & $T_{(1)}^2$         & $T_{(2)}^2$       & $T_{(3)}^2$ 
 \\ \hline\hline
 &     $bd$           & $\emptyset$ & $\parallel\neq$ & $1$, $2$, $3$ & $3$     \\
 $L_{1 \ldots 6}$  &   $b(\theta d)$    & $-6$        & $4$, $(R,R')$   & $1$, $2$, $3$ & $3$     \\
   &   $b(\theta^{2}d)$ & $|4|_{m}$   & $5$, $(S,S')$   & $\parallel=$  & $3$     
   \\ \hline\hline
   & $bd'$ & $|2|_m$ & 5, $(S,S')$ & $\parallel=$  & 3\\
  & $b(\theta d')$ & $\emptyset$ &  $\parallel\neq$  & 1,2,3 & 3\\
 $\ov{L}_{1\ldots 3}$  & $b(\theta^2 d')$ & $-3$ &  4, $(R,R')$  & 1,2,3 & 3
     \\ \hline\hline
 & $cd$ & $|2|_m$ & $(T,T')$ & $\parallel=$  & 3\\
$E_{1\ldots 3}$ & $c(\theta d)$ &  3 &  4 & 1,2,3  & 3\\
 & $c(\theta^2 d)$ & $\emptyset$ &  5 & 1,2,3 & 3
 \\\hline  
   \end{tabular}
\caption{Assignments of left- and right-handed leptons to a given intersection sector $x(\theta^k y)$ with counting of chiralities $\chi_{x(\theta^k y)}$ 
or multiplicities of the non-chiral matter content $\varphi_{x(\theta^k y)}$. The notation $E_i=(e_R^i,\nu_R^i)$ is taylored for the right-symmetric case
$USp(2)_c$ with $\sigma^2_c=0$. Family replication in the lepton sector is due to three intersection points 1,2,3 along $T^2_{(2)}$.}
\label{tab:leptons}
%
\end{minipage}
\begin{minipage}{74mm}
\setfloattype{table}
 \setlength{\tabcolsep}{1pt}
   \begin{tabular}{|c|c|c|c|c|c|}      \hline
\multicolumn{6}{|c|}{\text{\bf Multiplicities and localisations of quarks \& Higgses}}
\\ \hline\hline
 Particle &      $x(\theta^k y)$           &  $\chi/\varphi_{x(\theta^k y)}$   & $T_{(1)}^2$         & $T_{(2)}^2$       & $T_{(3)}^2$ 
 \\ \hline\hline
$Q_1,Q_2$ & $ab'$ & $-2$ & 4 & 1,4 & 3\\
$Q_3, Q_4$ & $a(\theta b')$ & $-2$ & 4,5 & 1 & 3 \\
$\ov{Q}$ & $a(\theta^2 b')$ & 1 & 5 & 1 & 3 
\\\hline\hline
$U_1,U_2$ & $ac$ &  2 & $\parallel=$  & 1,4 & 3\\
& $a(\theta c)$ & $\emptyset$ &  5 & 1 & 3\\
$U_3$ & $a(\theta^2 c)$ & 1 & 4 & 1 & 3
  \\ \hline\hline
$H_{1\ldots 6}$ & $bc$ & 6 & 4,5 & 1,2,3 & 3\\
$H_{7\ldots 9}$ & $b(\theta c)$ & 3 &  5 & 1,2,3 & 3\\
& $b(\theta^2 c)$ &   $|2|_m$ &  4  & $\parallel=$  & 3
\\\hline  
   \end{tabular}
\caption{Assignment and chirality $\chi_{x(\theta^k y)}$ or non-chiral counting $\varphi_{x(\theta^k y)}$ of left- and right-handed 
quarks and Higgses in the gauge enhanced phase $U(1)_c \to USp(2)_c$ with $U_i=(u_R^i,d_R^i)$ and $H_i=(H_u^i,H_d^i)$. While family
replication in the Higgs sector is related to the intersection points 1,2,3 on $T^2_{(2)}$, the quarks display 
a more intricate pattern on $T^2_{(1)} \times T^2_{(2)}$.}
\label{tab:quarks+Higgs}
%
%
\end{minipage}
\end{figure}
The situation is depicted in figure~\ref{fig:Leptons} for D6-branes $b$, $c$ and $(\theta d)$ supporting the left-handed leptons $L_i$, right-handed leptons
$E_j$ and six of the Higgs generations $H_k$.

\section{Yukawa couplings}

\subsection{Holomorphic three-point couplings}

\begin{figure}[ht!]
\sidecaption
\includegraphics[height=44mm]{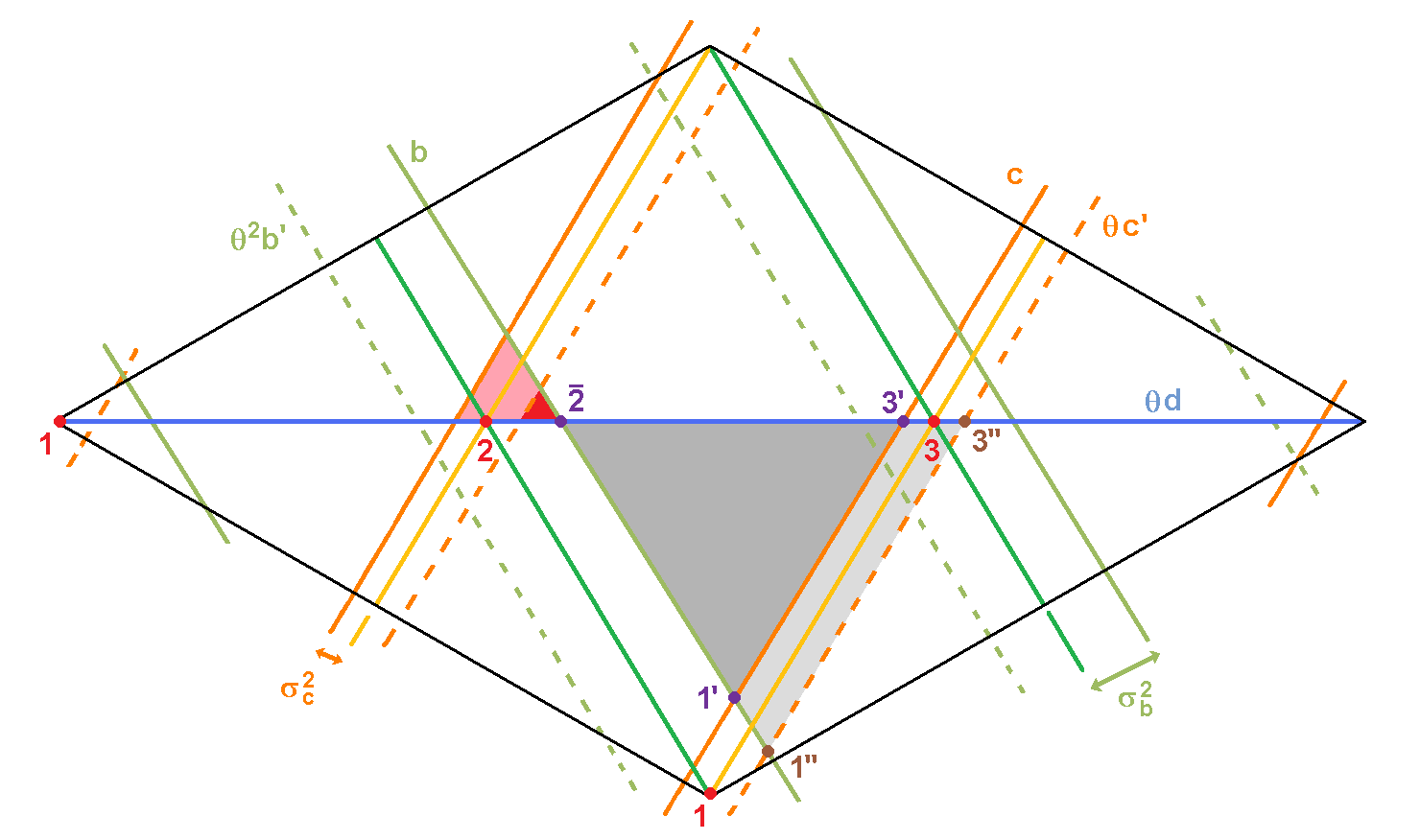}
\caption{Generation of Yukawa hierarchies within one generation (small red triangles) and 
in the flavour-mixing sectors (large grey triangles). In comparison to the original D6-brane 
configuration of figure~\protect\ref{fig:Leptons}, upon breaking $USp(2)_c \to U(1)_c$ by 
setting $\sigma^2_c \neq 0$, the coupling $e_R^2 L_2 H_{d}^2$ receives a suppression by a smaller area than the coupling
$\nu_R^2 L_2 H_{u}^2$ depicted in dark and dark+light red, respectively. The change of area
in the non-diagonal couplings acts in the opposite way as depicted in light+dark and dark grey, respectively, for the
$e_R^1L_2H_d^3$ and $\nu_R^1 L_2 H_u^3$ examples.
}
\label{fig:Displacements}
\end{figure}

Three-point couplings arise in perturbation theory from triangular worldsheets with edges given by the three D6-branes under which
matter is charged. For example, three chiral multiplets $\phi^i_{xy}$, $\phi^j_{yz}$ and $\phi^k_{zx}$
with respective charges $(\ov{\N}_x,\N_y)$, $(\ov{\N}_y,\N_z)$ and $(\ov{\N}_z,\N_x)$ under $U(N_x) \times U(N_y) \times U(N_z)$ lead 
to a superpotential term of the form
\begin{equation}
W= W_{ijk} \; \phi_{xy}^i \phi_{yz}^j \phi_{zx}^k
\qquad 
\text{ with }
\qquad 
W_{ijk} = \prod_{m=1}^3 \left(\sum_{{\cal A}_{ijk,(m)}} e^{-{\cal A}_{ijk,(m)}}\right)
.
\label{ThreePointInteraction}
\end{equation}
Besides the field theoretic consistency of charge neutrality, the string theoretic selection rule states
that $W_{ijk} \neq 0$ is only possible if $x$, $y$ and $z$ form a closed triangle with area ${\cal A}_{ijk,(m)}$
along $T^2_{(m)}$,  which can also be shrunken to a point or, for {\it fractional} or {\it rigid} D6-branes,
have one apex with vanishing angle. The infinite sums in~(\ref{ThreePointInteraction}) are due to the periodicity of the underlying six-torus. 
If e.g. only the two-torus $T^2_{(2)}$ is considered with vanishing Wilson lines $\tau^2_x \equiv 0$ for all D6-branes $x \in \{a,b,c,d,h\}$, 
the sum over areas can be compactly written 
as~\cite{Cremades:2003qj,Cremades:2004wa}, 
\begin{equation}
\vartheta \targ{\delta_2}{0}(\frac{t_2\mathcal{A}_{(2)}}{2\pi}) = \sum_{l \in \Z} e^{-t_2\mathcal{A}_{(2)} (\delta_2 + l)^2}
\quad \stackrel{\delta_2=0,t_2=3}{=} \quad  1 + 2 \, e^{-3 {\cal A}_{(2)}} + 2 \, e^{-12 {\cal A}_{(2)}} + \ldots
,
\label{Eq:Worldsheet-theta}
\end{equation}
with $t_2 \in \mathbb{N}$ the (absolute value of the) product of torus intersection numbers, $\delta_2$ a linear function of the 
location of intersections and the corresponding displacements $\sigma^2_x$ and ${\cal A}_{(2)}>1$ the 
two-torus volume in units of $2\pi\alpha'$. The sum in~(\ref{Eq:Worldsheet-theta}) is dominated by the term $l=0$,
as can be seen in the example of the family diagonal lepton Yukawa couplings at $b$, $c$ and $(\theta d)$ intersections 
with $\delta_2=0$ and $t_2=3$. The sum on the r.h.s. of~(\ref{Eq:Worldsheet-theta})
converges very fast since already the first non-trivial contribution $2 e^{-3 {\cal A}_{(2)}}< 2 e^{-3}<0.1$ 
is tiny for areas ${\cal A}_{(2)} > 1$ in the geometric regime of type IIA string compactifications.
Non-diagonal matter couplings arise typically from worldsheets with a non-vanishing triangular area, 
which can be parameterised as a fraction $\delta_2 \in \mathbb{Q}$ of the two-torus area ${\cal A}_{(2)}$. 
Typical numerical examples of such suppression factors are given in table~\ref{tab:Suppressions}.
\begin{table}[ht!]
\tabsidecaption
 \setlength{\tabcolsep}{2pt}
   \begin{tabular}{|c||c|c|c|c|c|c|c|c|c|}      \hline
\multicolumn{10}{|c|}{\text{\bf Suppression factors of pert. 3-point couplings by  worldsheet areas}}
\\ \hline\hline
${\cal A}$ & 1 & 2 & 3 & 5 & 10 & 20 & 30 & 50 & 100
\\\hline\hline
$e^{-{\cal A}/12}$ & 0.920 & 0.846 & 0.779 &  0.659 & 0.435 & 0.189 & 0.082 & 0.016 & $2 \cdot 10^{-4}$ 
\\
$e^{-{\cal A}/6}$ & 0.846 & 0.717 & 0.607 & 0.435 & 0.189 & 0.036 & 0.007 & $2 \cdot 10^{-4}$ & $6 \cdot 10^{-8}$  
\\
$e^{-{\cal A}/4}$ & 0.779 & 0.607 & 0.472 & 0.287 & 0.082 & 0.007 & $6 \cdot 10^{-4}$ & $4 \cdot 10^{-6}$ & $10^{-11}$
\\
$e^{-{\cal A}/3}$ & 0.717 & 0.513 & 0.368 & 0.189 & 0.036 & 0.001 & $5 \cdot 10^{-5}$ &  $6 \cdot 10^{-8}$ &  $3 \cdot 10^{-15}$
\\\hline
${\cal A}^{-1/4}$ & 1 & 0.841 & 0.760 & 0.669 & 0.562 & 0.473 & 0.427 & 0.376 & 0.316
  \\\hline
   \end{tabular}
\caption{Numerical values of area suppressed Yukawa couplings for various sizes of the two-torus area ${\cal A}$
in units of $2\pi\alpha'$. The larger the area, the better the approximation by the first contribution to the family diagonal
couplings.}
\label{tab:Suppressions}
\end{table}
A hierarchical structure of  Yukawa couplings for lepton and quark families can thus be generated by a suitable choice of matter localisations and corresponding 
triangular areas. Two ways of tuning arise by varying the ratios ${\cal A}_{(i)}/{\cal A}_{(j)}$ of two-torus volumes $(i \neq j)$
and by choosing different continuous displacement
parameters $\sigma^2_x$ for each D6-brane $x$ along $T^2_{(2)}$ as displayed in figure~\ref{fig:Displacements}.
%

\subsection{Physical Yukawa couplings}

The couplings in the holomorphic superpotential~(\ref{ThreePointInteraction}) are in the supergravity theory dressed by a non-holomorphic
prefactor involving the K\"ahler potential ${\cal K}$ and its derivatives, the K\"ahler metrics $K_{xy}$,
\begin{equation}\label{Eq:Physical_Yukawas}
Y_{ijk} = \left( K_{xy} \, K_{yz} \, K_{zx} \right)^{-1/2} \; e^{\kappa_4^2 \; {\cal K}/2} \; W_{ijk}\;
.
\end{equation}
While the K\"ahler potential ${\cal K}$ in the exponential is universal for all three-point interactions, the K\"ahler metrics
$K_{\mathbf{R}_a} = \bigl(f(S,U)/\prod_{i=1}^3\sqrt{{\cal A}_{(i)}}\bigr) F(\vec{\phi}_{ab})$
with  $f(S,U)=(S\, \prod_{i=1}^{h_{21}^{\rm bulk}} U_i)^{-\alpha/4}$ and $(h_{21},\alpha)=(1,2)$ for $T^6/\Z_6'$
depend on the intersection angles of the corresponding D6-branes through $F(\vec{\phi}_{ab})= \prod_{i=1}^3 \sqrt{
\Gamma(|\phi^{(i)}_{ab}|) / \Gamma(1-|\phi^{(i)}_{ab}|) }^{\; -\sgn(\phi^{(i)}_{ab})/\sgn(I_{ab}) }$
or $F(0^{(i)},\phi^{(j)}_{ab},\phi^{(k)}_{ab})=\sqrt{2\pi {\cal A}_{(i)} V_{ab}^{(i)}}$ and $V_{ab}^{(i)}$ the normalised one-cycle volume 
of D6-branes $a$, $b$  parallel along $T^2_{(i)}$~\cite{Honecker:2011sm,Honecker:2011hm}.
The non-holomorphic prefactor in~(\ref{Eq:Physical_Yukawas}) provides thus a second potential source for the generation of Yukawa hierarchies
through the functions $F(\vec{\phi}_{ab})$~\cite{Honecker:2012jd}, 
namely either different choices of supersymmetric angle configurations $(\vec{\phi}_{ab})$
or the normalised one-cycle volume $V_{ab}^{(i)}$ times two-torus area ${\cal A}_{(i)}$.

\subsection{Yukawa couplings for the Standard Model with $USp(6)_h$ on $T^6/\Z_6'$}

The holomorphic Yukawa interactions for the D6-brane configuration in table~\ref{RepresentationsSMOnBranes} can be derived from the localisations of SM
matter states in table~\ref{tab:leptons}. Since all left-handed leptons $L_i$ arise at $b(\theta d)$ intersections and all right-handed leptons 
$E_j$ at $c(\theta d)$ intersections, closed triangular worldsheets are only spanned if the third apex is formed by one of the $bc$ intersections, 
at which six of the Higgs families $H_k$ are localised.
The dominant terms provide diagonal Yukawa couplings with one Higgs generation
per lepton family, whereas all area suppressed couplings providing lepton flavour mixing couplings,
\begin{equation}\nonumber
\begin{aligned}
\left.\begin{array}{r}
W_{E_{i}L_{i}H_{i}} \sim {\cal O}(1)
\\
W_{E_{i}L_{i+3}H_{i+3}} \sim {\cal O}(e^{-\mathcal{A}_{(1)}/12})
\\
W_{E_{i}L_{i+3}H_{i+3}} \sim {\cal O}(e^{-\mathcal{A}_{(1)}/12})
\end{array} \right\} &
 \qquad
\text{with}
\qquad
i=1,2,3
,
\\
W_{E_{i}L_{j}H_{k}} \sim {\cal O}(e^{-\mathcal{A}_{(2)}/6})
&
\qquad
\text{with}
\qquad
(i,j,k)
\quad
\text{permutations of }
(1,2,3)
,
\end{aligned}
\end{equation}
%
\begin{equation}\nonumber
\begin{aligned}
\left.\begin{array}{l}
W_{E_{i}L_{j}H_{k}} \sim {\cal O}(e^{-\mathcal{A}_{(1)}/12-\mathcal{A}_{(2)}/6})\\
W_{E_{i}L_{j}H_{k-3}} \sim {\cal O}(e^{-\mathcal{A}_{(1)}/3-\mathcal{A}_{(2)}/6})
\end{array} \right\} 
& \qquad
\text{with}
\qquad
\left\{\begin{array}{rr}
i=1 & (j,k)=(5,6),(6,5) \\
2 & (6,4),(4,6)\\
3 & (4,5),(5,4)
\end{array}\right.
.
\end{aligned}
\end{equation}
Since all worldsheets are spanned by triangles with edges $b$, $c$ and $(\theta d)$, the non-holomorphic prefactor
$(K_{bc} \, K_{c(\theta d)} \, K_{b(\theta d)} )^{-1/2} =  g(S,U,{\cal A})  \frac{1}{2(50)^{1/4}}$ with $g(S,U,{\cal A})\equiv\frac{({\cal A}_{(1)}{\cal A}_{(2)}{\cal A}_{(3)})^{3/4}}{f(S,U)^{3/2}}$ and
$\frac{1}{2(50)^{1/4}} \approx  0.188$ is universal in the lepton sector. 

The quark Yukawa couplings display a different pattern since left-and right handed quarks arise each in two sectors,  $ab'$ and $a(\theta b')$ 
and $ac$ and $a(\theta^2 c)$, respectively. The $ac$ sector is examplary for the exceptional situation that on {\it fractional} D6-branes
chiral matter states can arise at one vanishing intersection angle along $T^2_{(1)}$. We therefore include the option of a triangle with zero angle and vanishing area,
for which the infinite sum formula~(\ref{Eq:Worldsheet-theta}) may require modification. The dominant Yukawa interactions
involve only two right- and three left-handed quark generations $U_1,U_3$ and $Q_1, Q_3, Q_4$ plus three Higgs generations $H_1, H_4, H_7$, 
one of which also couples to the first lepton generation,
\begin{equation}\nonumber
\begin{aligned}
W_{U_i Q_j H_k} \sim \left\{\begin{array}{rr} {\cal O}(1) & \text{with} \quad
(i,j,k)=(3,1,1),(3,4,7),(1,3,1),(1,4,4) ,
\\
{\cal O}(e^{-\mathcal{A}_{(1)}/4}) & (3,1,4), (3,3,7),
\\
{\cal O}(e^{-\mathcal{A}_{(2)}/12}) &  (3,2,l), (2,3,5-l),(2,4,8-l),
\\
{\cal O}(e^{-\mathcal{A}_{(2)}/6}) & (3,4,6+l),
\\
{\cal O}(e^{-\mathcal{A}_{(1)}/4-\mathcal{A}_{(2)}/12}) & (3,2,3+l),
\\
{\cal O}(e^{-\mathcal{A}_{(1)}/4-\mathcal{A}_{(2)}/6}) & (3,3,6+l)
.
\end{array}\right.
\text{with }
l \in \{2,3\}.
\end{aligned}
\end{equation}
Since quark Yukawa couplings arise from three different types of triangular worldsheets, distinct non-holomorphic prefactors arise,
$(K_{ab'} \,K_{a(\theta^2 c)} \,  K_{bc} )^{-1/2} =g(S,U,{\cal A}) \frac{1}{10^{3/4}}$ with $ \frac{1}{10^{3/4}}\approx 0.178$, \linebreak
$(K_{a(\theta b')} \,K_{a(\theta^2 c)}  \, K_{b(\theta^2 c)} )^{-1/2} =g(S,U,{\cal A}) \frac{1}{5 \cdot 2^{1/4}}$ with $\frac{1}{5 \cdot 2^{1/4}}\approx  0.168$
and $(K_{a(\theta b')} \,K_{ac} \, K_{b(\theta^2 c)}  )^{-1/2} =g(S,U,{\cal A}) \frac{3^{1/8}}{2(25\pi v_1)^{1/4}}$
with $\frac{3^{1/8}}{2(25\pi {\cal A}_{(1)})^{1/4}}\approx \frac{0.193}{{\cal A}_{(1)}^{1/4}}$, which, however, are numerically all of the same order of magnitude as the 
one for the leptons, cf. also the last line of table~\ref{tab:Suppressions} for the very mild additional suppression by ${\cal A}^{-1/4}$
in the last case.

\subsection{Masses for vector-like matter}

The vector-like matter spectrum can be decomposed into three types with different string theoretic
origin and consequently distinct mechanisms by which they acquire masses~\cite{Honecker:2012jd}. 
The first kind consists of ${\cal N}=2$ supersymmetric sectors with two D6-branes parallel along $T^2_{(2)}$,
for which masses are provided by separations of the D6-branes. The matter states in brackets on the second 
and third line of  equation~(\ref{FullSpectrum2}) are of this type with $\sigma^2_b \neq 0$ and $\sigma^2_c \neq 0$,
respectively, providing the masses. 

A second type of vector-like matter arises at two distinct ${\cal N}=1$
supersymmetric intersections such as the left-handed quarks $Q_i$ at $a(\theta^k b')_{k \in \{0,1\}}$ and 
the conjugate $\ov{Q}$ at an $a(\theta^2 b')$ intersection in table~\ref{tab:quarks+Higgs}. The masses arise via 
couplings to SM singlets, e.g. those inside the adjoint representations $A_i$ of $U(3)_a$, 
$({\bf 9}_{U(3)_a}) = ({\bf 8}_{SU(3)_a})_0 + (\1)_0$, 
\begin{equation}
\begin{aligned}
W_{\ov{Q}Q_i A_2} \sim \left\{\begin{array}{rr}
{\cal O}(1) & i=4
\\
(e^{-\mathcal{A}_{(1)}/8}) & 1
\\
(e^{-\mathcal{A}_{(1)}/4}) & 3
\\
(e^{-\mathcal{A}_{(1)}/8-\mathcal{A}_{(2)}/4}) & 2
\end{array}\right.
,
\qquad 
W_{\ov{W}_{\ov{i}} W_j A_2} \sim \left\{\begin{array}{rr}
{\cal O}(1) & \ov{i},j\in \{2,3\}
\\
(e^{-\mathcal{A}_{(1)}/8}) & (\ov{i},j)=(1,1)
\end{array}\right.
,
\end{aligned}
\end{equation}
which provide mass terms if $A_2$ receives a {\it vev} along some flat direction of the 
(trivial D-term) and F-term potential,
see~\cite{Honecker:2012jd} for more details.
The example on the r.h.s. shows the same mechanism for the $(\ov{\3}_{\Anti},\1)_{1/3}$ representations $W_j$
and their conjugates $\ov{W}_{\ov{i}}$. By combining several {\it vev}s of singlets inside the adjoints of 
$U(3)_a$ and $U(2)_b$, all states on the first line and the remaining charged states on the last line 
of~(\ref{FullSpectrum2}) are rendered massive. The matter states with $SU(2)_b \times U(1)_Y$ charge in the 
`hidden' spectrum~(\ref{FullSpectrum3}) acquire masses by a similar mechanism, namely if the $({\bf 15}_{\Anti})_0$ of 
$USp(6)_h$ receives some {\it vev}.

The last type  of vector-like matter consists of three lepton $(L_{3+i}, \ov{L}_i)$ and nine Higgs $(H_u^i,H_d^i)$
pairs in equation~(\ref{FullSpectrum1}), which are `chiral' w.r.t. the anomalous $U(1)_b$ symmetry. 
These states couple perturbatively to the symmetric representations $\ov{b}_i$ of $U(2)_b$ or its conjugate $b_i$
instead of the adjoints for the above mentioned vector-like quark pairs, e.g. the non-suppressed couplings read
$W_{\ov{L}_i b_i L_{3+i}} , W_{H_i \ov{b}_i H_j} \sim {\cal O}(1)$.
Any {\it vev} will then lead to a gauge symmetry breaking $SU(2)_b \to U(1)_{I_3}$, which might happen
at an intermediary energy scale between $M_{\text{weak}}$ and $M_{\text{string}}$~\cite{Honecker:2012jd}. 
Alternatively, D-brane instantons or higher order $n$-point couplings might provide mass terms without 
gauge symmetry breaking.

\section{Conclusions}

We presented an estimation of the relative order of magnitude of Yukawa couplings in the lepton and quark sector for the SM on $T^6/\Z_6'$
based on a scrutiny of the worldsheets with minimal areas. 
While the dominant terms for the leptons are flavour diagonal, mixings already occur at leading order in the quark sector.
Our derivation of perturbative three-point couplings, which also includes masses for the vector-like matter states
through Higgs-like couplings to Standard Model singlets, relies on methods developed for the six-torus.
A thorough string theoretic derivation is needed to clarify if further stringy selection rules such as discrete symmetries restrict the sums over worldsheet areas 
and if a similar lattice sum  occurs as well for the special case of one vanishing angle. Last but not least, higher order and D-instanton couplings
are expected to modify the subleading behaviour.


\begin{acknowledgement}
 The work of G.H. is partly supported by the  ``Research Center Elementary Forces and Mathematical Foundations'' (EMG)
at the Johannes Gutenberg-Universit\"at Mainz.
The research of J.V. has been supported in part by the Belgian Federal Science Policy Office through the Interuniversity Attraction Pole 
IAP VI/11 and by FWO-Vlaanderen through project G011410N.  
\end{acknowledgement}



\begin{thebibliography}{10}

\bibitem{Gmeiner:2005vz}
  F.~Gmeiner, R.~Blumenhagen, G.~Honecker, D.~L{\"u}st and T.~Weigand,
  JHEP {\bf 0601} (2006) 004
%
\bibitem{Gmeiner:2007we}
  F.~Gmeiner, D.~L{\"u}st and M.~Stein,
  JHEP {\bf 0705} (2007) 018
%
\bibitem{Gmeiner:2008xq}
  F.~Gmeiner and G.~Honecker,
  JHEP {\bf 0807} (2008) 052
%
\bibitem{Lust:2003ky}
  D.~L{\"u}st and S.~Stieberger,
  Fortsch.\ Phys.\  {\bf 55} (2007) 427
%
\bibitem{Akerblom:2007np}
  N.~Akerblom, R.~Blumenhagen, D.~L{\"u}st and M.~Schmidt-Sommerfeld,
  Phys.\ Lett.\ B {\bf 652} (2007) 53
%
\bibitem{Blumenhagen:2007ip}
  R.~Blumenhagen and M.~Schmidt-Sommerfeld,
  JHEP {\bf 0712} (2007) 072
%
\bibitem{Gmeiner:2009fb}
  F.~Gmeiner and G.~Honecker,
  Nucl.\ Phys.\ B {\bf 829} (2010) 225
%
\bibitem{Cremades:2003qj}
  D.~Cremades, L.~E.~Ib{\'a}{\~n}ez and F.~Marchesano,
  JHEP {\bf 0307} (2003) 038
%
\bibitem{Cremades:2004wa}
  D.~Cremades, L.~E.~Ib{\'a}{\~n}ez and F.~Marchesano,
  JHEP {\bf 0405} (2004) 079
%
\bibitem{Bailin:2006zf}
  D.~Bailin and A.~Love,
  Nucl.\ Phys.\ B {\bf 755} (2006) 79
   [Nucl.\ Phys.\ B {\bf 783} (2007) 176]
%
\bibitem{Bailin:2007va}
  D.~Bailin and A.~Love,
  Phys.\ Lett.\ B {\bf 651} (2007) 324
   [Erratum-ibid.\ B {\bf 658} (2008) 292]
%
\bibitem{Gmeiner:2007zz}
  F.~Gmeiner and G.~Honecker,
  JHEP {\bf 0709} (2007) 128
%
\bibitem{Bailin:2008xx}
  D.~Bailin and A.~Love,
  Nucl.\ Phys.\ B {\bf 809} (2009) 64
%
\bibitem{Blumenhagen:2002gw}
  R.~Blumenhagen, L.~G{\"o}rlich and T.~Ott,
  JHEP {\bf 0301} (2003) 021
%
\bibitem{Honecker:2004kb}
  G.~Honecker and T.~Ott,
  Phys.\ Rev.\ D {\bf 70} (2004) 126010
   [Erratum-ibid.\ D {\bf 71} (2005) 069902]
%
\bibitem{Honecker:2004np}
  G.~Honecker,
  Mod.\ Phys.\ Lett.\ A {\bf 19} (2004) 1863
%
\bibitem{Blumenhagen:2005tn}
  R.~Blumenhagen, M.~Cveti\v{c}, F.~Marchesano and G.~Shiu,
  JHEP {\bf 0503} (2005) 050
%
\bibitem{Forste:2010gw}
  S.~F{\"o}rste and G.~Honecker,
  JHEP {\bf 1101} (2011) 091
%
\bibitem{Honecker:2012jd}
  G.~Honecker and J.~Vanhoof,
  arXiv:1201.3604 [hep-th].
%
\bibitem{Honecker:2011sm}
  G.~Honecker,
  arXiv:1109.3192 [hep-th].
%
\bibitem{Honecker:2011hm}
  G.~Honecker,
  arXiv:1109.6533 [hep-th].





\end{thebibliography}
\end{document}